\documentclass[sigconf]{acmart}
\AtBeginDocument{%
  \providecommand\BibTeX{{%
    \normalfont B\kern-0.5em{\scshape i\kern-0.25em b}\kern-0.8em\TeX}}}
\copyrightyear{2022} 
\acmYear{2022} 
\setcopyright{acmcopyright}
\acmConference[MM '22]{Proceedings of the 30th ACM International Conference on Multimedia}{October 10--14, 2022}{Lisboa, Portugal} 
\acmBooktitle{Proceedings of the 30th ACM International Conference on Multimedia (MM '22), October 10--14, 2022, Lisboa, Portugal} 
\acmPrice{15.00} 
\acmDOI{10.1145/3503161.3548769} \acmISBN{978-1-4503-9203-7/22/10}
\usepackage{xspace}

\def\persrec{\emph{PersiC}\xspace}

\author{Qi Yang}
\email{yangqi@itmo.ru}
\affiliation{%
   \institution{ITMO Univerity}
   \city{Saint Petersburg}
   \country{Russia}}

\author{Sergey Nikolenko}
\email{sergey@logic.pdmi.ras.ru}
\affiliation{%
   \institution{Steklov Institute of Mathematics at St. Petersburg}
   \city{Saint Petersburg}
   \country{Russia}}

\author{Alfred Huang}
\email{alfred@somin.ai}
\affiliation{%
\institution{Multimedia Research Lab, SoMin.ai}
\country{Singapore}}

\author{Aleksandr Farseev}
\email{farseev@itmo.ru}
\affiliation{%
   \institution{ITMO Univerity}
   \city{Saint Petersburg}
   \country{Russia}}


\begin{document}

\title{Personality-Driven Social Multimedia Content Recommendation}

\begin{abstract}
Social media marketing plays a vital role in promoting brand and product values to wide audiences. In order to boost their advertising revenues, global media buying platforms such as \emph{Facebook Ads} constantly reduce the reach of branded organic posts, pushing brands to spend more on paid media ads. In order to run organic and paid social media marketing efficiently, it is necessary to understand the audience, tailoring the content to fit their interests and online behaviours, which is impossible to do manually at a large scale. At the same time, various personality type categorization schemes such as the Myers-Briggs Personality Type indicator make it possible to reveal the dependencies between personality traits and user content preferences on a wider scale by categorizing audience behaviours in a unified and structured manner. Still, McKinsey-style manual categorization is a very labour-intensive task that is probably impractical in a real-world scenario, so automated incorporation of audience behaviour and personality mining into industrial applications is necessary. This problem is yet to be studied in depth by the research community, while the level of impact of different personality traits on content recommendation accuracy has not been widely utilised and comprehensively evaluated so far. Even worse, there is no dataset available for the research community to serve as a benchmark and drive further research in this direction. The present study is one of the first attempts to bridge this important industrial gap, contributing not just a novel personality-driven content recommendation approach and dataset, but also facilitating a real-world ready solution which is scalable and sufficiently accurate to be applied in real-world settings. Specifically, in this work we investigate the impact of human personality traits on the content recommendation model by applying a novel personality-driven multi-view content recommender system called \emph{Personality Content Marketing Recommender Engine}, or \persrec. Our experimental results and real-world case study demonstrate not just \persrec's ability to perform efficient human personality-driven multi-view content recommendation, but also allow for actionable digital ad strategy recommendations, which when deployed are able to improve digital advertising efficiency by over 420\% as compared to the original human-guided approach.
\end{abstract}


\maketitle

\section{Introduction}

Over the past decade, social networks have become an integral part of our lives. Hundreds of millions, if not billions, of people, use social networks on a daily basis, reading their feeds, communicating with friends, watching videos... and watching ads. The growing use of social media has led to a corresponding exponential growth in social media advertising, with ads on \emph{Facebook} and other social media playing an increasingly important role in product promotion and customer human making. At the same time, another important recent trend to note is a decisive \emph{decrease} of the so-called ~\cite{organic} audience reach, which is the number of people who can see one's post on social media with no paid digital advertising  (also known as Facebook posts boosting) involved. It looks like the users are becoming increasingly harder to ``trick'' into engaging with the digital content and spreading the word about the ads they see organically, while Facebook itself reduces the organic ad impressions artificially leading brands to scale on digital via paid ads.

\begin{figure}[!t]\centering
     \includegraphics[width=\linewidth]{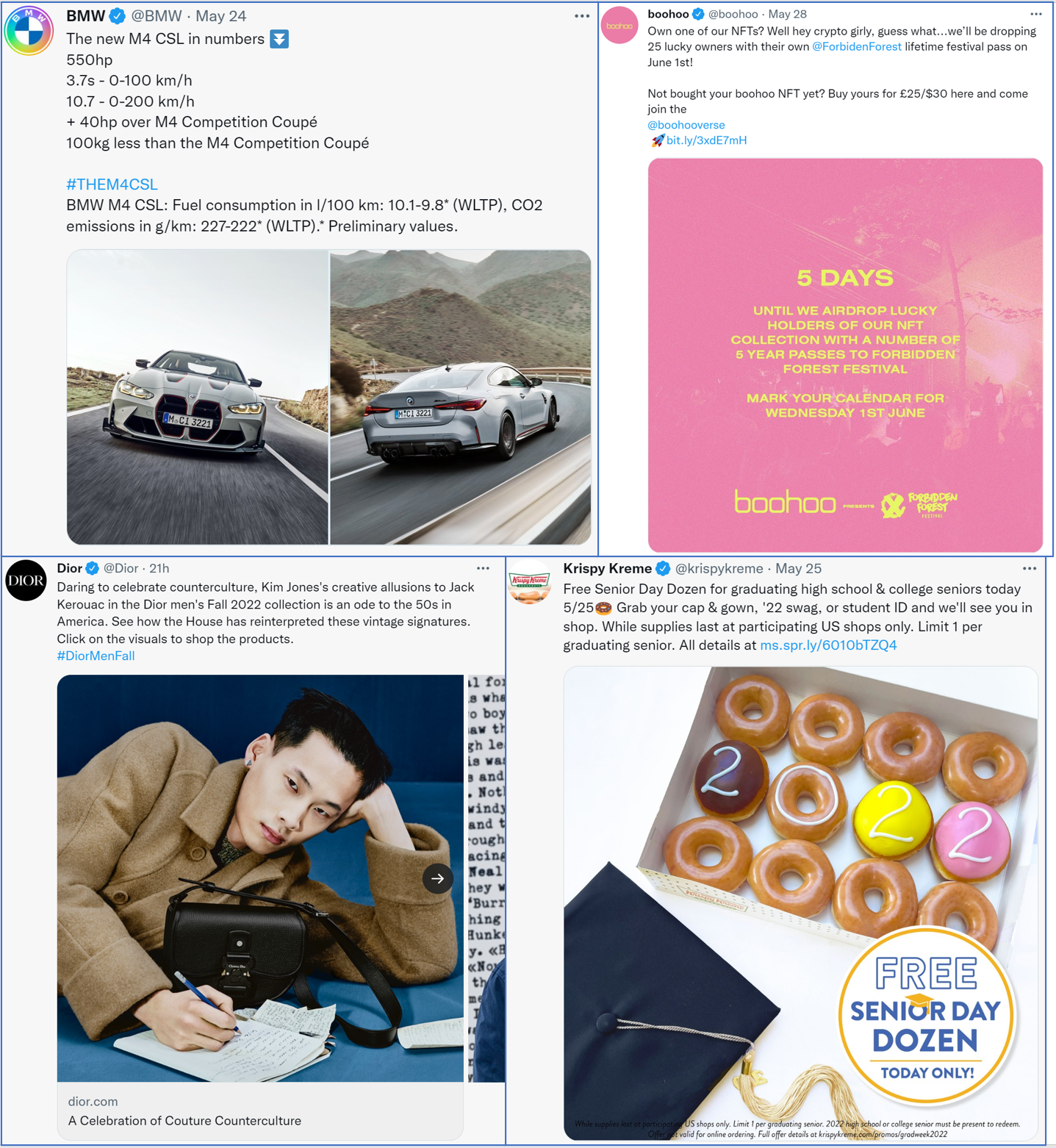}

     \caption{Four sample brand posts from BMW, BOOHOO, \emph{Dior}, and \emph{KrispyCreme} on \emph{Twitter}; note the different content styles brands use to promote their products or events.}
     \label{fig:vbrand}
\end{figure}

Following this global trend, brands and marketing agencies promote products on social media via paid advertising campaigns on ad distribution engines such as \emph{Meta Ads} (formerly \emph{Facebook Ads}). A typical campaign consists of multiple ads grouped together, and even in the simplest case an ad can contain both text and image data. Moreover, modern advertisers and agencies, like Somin.ai, are often constrained by particular ad targeting capabilities aimed at capturing the desired social network audience base (e.g., casual fitness practitioners), which, in turn, is expected to be associated with higher ad conversion rates (e.g., more likely to purchase sports-related products). Figure~\ref{fig:vbrand} shows four sample brand posts that illustrate how brands use different content styles to promote their products.


In order to find a viable solution to this important problem for the content marketing industry, many recommender systems have been proposed in both academia and industry. In addition to classical collaborative filtering, modern recommender systems are able to leverage multi-view item content (in our case, text and images), user demographics, temporal criteria, and many more (see Section~\ref{sec:related}). In this work, we are making another step towards personalized content recommendation by proposing to leverage, arguably, the most influential human decision-making component: our \emph{personality}. Specifically, we use \emph{personality traits}, generalized labels proposed in psychology to classify mental activity patterns related to acquiring information, making decisions, and generally dealing with life; it stands to reason that such labels can be utilized to understand content preferences and tailor ads to a specific person.

In this work, we have adopted a widely-utilized personality categorization scheme called the \emph{Myers–Briggs Type Indicator} (MBTI), which breaks down human personality into 16 types, one for every composition of four binary components (traits): \textbf{E}xtroversion vs. \textbf{I}ntroversion, \textbf{S}ensing vs. i\textbf{N}tuition, \textbf{T}hinking vs. \textbf{F}eeling, and \textbf{J}udging vs. \textbf{P}erceiving. MBTI categorization is widely accepted in psychology and used for personality profiling in, e.g., career planning and market research. However, until now there has not been much research devoted to the incorporation of personality traits into industry-applicable actionable content recommendations, and we are making one of the first large-scale attempts to solve the problem at a new level of adaptability and performance.











Encouraged by the industrial needs and research gap described above, we pose the following research questions:
\begin{itemize}
    \item[RQ1] is it possible to predict personality traits from data available to recommender systems in a social network setting?
    \item[RQ2] can we improve content recommendations by incorporating personality traits predicted from social media data?
    \item[RQ3] what is the difference in content consumption preferences between users exhibiting different personality traits and if such insights could add value to digital advertising routines?
\end{itemize}

To answer our research questions, in this study we make one of the first attempts to perform personality-driven content recommendation, that is sufficiently accurate to be applied in real-world settings.  Particularly, we investigate the impact of human personality on the content recommendation model performance. Our experimental results and real-world case study demonstrate not just \persrec's ability to perform efficient human personality-driven multi-view content recommendation, but also allow for gaining actionable digital ad strategy insights, which are able to drive 420\% digital advertising performance boost as compared to traditional human-based approaches\footnote{Special thanks to Somin.ai for providing the environment of experiment. }.
The paper is organized as follows: in Section~\ref{sec:related} we give an overview of previous research on personality traits and recommender systems, Section~\ref{sec:data} introduces the datasets used to learn a personality trait model and give recommendations, Section~\ref{sec:method} presents our approach, the \persrec framework for recommendation with personality trait features, Section~\ref{sec:eval} presents a comprehensive experimental evaluation, and Section~\ref{sec:concl} concludes the paper.

\section{Related work}\label{sec:related}

This work presents a new recommender system with personality features so the nearest prior art is related to recommendation systems. Classical collaborative filtering, i.e., completion of the user-item feedback matrix, is mostly based on matrix factorization (MF)~\citep{mnih2007probabilistic,bell2007modeling,reference/rsh/KorenB11}; MF-based approaches still present a competitive baseline, and many variations and applications have been developed over the years~\cite{RecSysHandbook}, but there are several viable alternatives even for the standard matrix completion problem. In particular, we note \emph{factorization machines}~\cite{5694074} that combine SVMs and matrix factorization and are able to train on implicit feedback. Most popular modern approaches utilize user and item \emph{embeddings}, vectors in a space that should represent the similarity between users and items that would be useful for recommendations, and the mapping to the embedding space does not have to be linear as in classical MF. In particular, autoencoder-based methods replace direct learning of the user feature matrix with learning a function that maps user feedback to user embeddings; this field began with shallow autoencoders~\citep{sedhain2015autorec,wu2016collaborative} and then moved to variational autoencoders that allow training deeper and more complex models~\citep{liang2018variational,lobel2019ract,kim2019enhancing,shenbin2020recvae,mirvakhabova2020performance,farseev2017cross}. Another approach to collaborative filtering is based on graph convolutional networks, including NFCF~\citep{wang2019neural} and LightGCN~\citep{he2020lightgcn}; training such networks is computationally intensive but they demonstrate impressive performance, and recent approaches such as GF-CF~\citep{shen2021powerful} and UltraGCN~\citep{mao2021ultragcn} improved both performance and computational efficiency. As collaborative filtering baselines, in this work we use basic MF, \emph{Neural Collaborative Filtering} (NeuCF) that uses deep neural networks to learn user and item embeddings~\cite{10.1145/3038912.3052569}, and \emph{Bilateral Variational Autoencoder for Collaborative Filtering} (BiVAECF) that learns embeddings via variational autoencoders~\cite{10.1145/3437963.3441759}.

We propose a model that takes into account personality traits, so we need recommender systems able to incorporate external information. Such recommender systems usually employ a content analysis module that extracts item representations from unstructured information such as text and images and a user profiling module that extracts user representations from features such as demographics (or, in our case, personality traits). Most modern approaches are actually \emph{hybrid} recommender systems~\cite{DBLP:journals/corr/abs-1901-03888,farseev2015cross} able to leverage both content/demographics and the user-item matrix; additional information can include external knowledge bases~\cite{10.1109/GreenCom-CPSCom.2010.144}, content-related features such as tags or keywords~\cite{6689996}, dynamically chosen expert users whose opinions should be trusted more~\cite{6762976}, and so on. In embedding-based approaches, additional user and item features are used to inform either the neural network producing these embeddings or the neural network that performs recommendations based on them; such approaches include the widely used DeepCoNN~\cite{10.1145/3018661.3018665}, \emph{YouTube} recommendations~\cite{10.1145/2959100.2959190}, embeddings based on topic models~\cite{DBLP:journals/corr/WangWY14}, and more. As the primary baseline for this work, we selected the Personalized Content Discovery (PCD) model~\cite{Francesco} because it was tailored to a similar problem of content discovery for brands but also note several other works that extend recommender systems with extra features and extra data modalities~\cite{10.1007/978-981-16-6372-7_25,10.1145/3447683,10.1145/3366423.3380002,10.1145/3459637.3482136,10.1145/2939672.2939673, farseev2018somin,9428193,liu2022hs,9811387,9138776,9726844}.

Finally, we note prior art in the research on personality traits. There have been several research directions that model human personality traits with different statistical approaches. First, the \emph{Big Five}, or NEO-PI model proposed in~\cite{Big5} was based on the assumption that human personality is reflected in their written language, so statistical analysis of the latter can inform us of the former. Inspired by this idea, the LIWC word categorization scheme~\cite{Pennebaker1999LinguisticSL} later provided a numerical connection between personality traits and written language utilization patterns; relations between the Big Five and MBTI personality types are well known~\cite{FURNHAM1996303}.

Automatic personality profiling started in the early 2000s, with cross-disciplinary research using machine learning techniques for automatic human personality inference based on data collected from questionnaires and personality tests~\citep{Mairesse, Argamon05}. Early studies, however, were conducted on relatively small datasets and did not make use of the huge data provided by social networks. This problem was acknowledged and partially mitigated in the \emph{MyPersonality} project~\citep{Mypersonality} that was able to provide the first large-scale personality-labeled dataset that includes user-generated data from \emph{Facebook}. This dataset soon became the basis for larger-scale studies in social media personality profiling research~\citep{attention1,attention2,attention3}. These studies made a big leap in the field, but most of them were still focused on a single data source such as, e.g., \emph{Facebook}, or a single data modality, e.g., text. In particular, the Linguistic Inquiry and Word Count (LIWC) works are mostly focused on text processing and predicting personality by using personality-labeled word categories~\citep{liwc1,liwc2}, while the works~\cite{glove1,glove2} instead utilized pretrained \emph{GloVe} embeddings~\citep{glove} and were the first to report the results of machine learning-driven unimodal personality inference. Image-based personality recognition methods can utilize, e.g., correlating specific image features with personality traits~\cite{KIM20181101,Qi_Aleksandr_Andrey_2020}, localization of attention on images a user liked~\cite{10.1007/s11063-019-09987-7}, joint learning of personality types and emotions from facial images~\cite{8897617}, and more. Finally, there are user behavior-based personality extraction methods based, e.g., on \emph{Facebook} likes~\cite{attention3}.

Modern social network data, however, is multi-source, multi-view, and multimodal, combining text, images, videos, and other data sources for a single person. There exist several studies that have approached user profiling from a multimodal data perspective. For example,~\citet{farseev2015harvesting} proposed a multimodal ensemble model for the demographic profiling problem from multimodal data, a work that was later extended to leverage sensor data and multi-source multi-task learning for wellness profiling~\citet{farseev2017tweetCanBeFit,farseev2017tweetFit}. In~\citet{buraya2017towards}, relationship status between social network users was predicted by applying classical machine learning techniques on early-fused data from \emph{Twitter}, \emph{Instagram}, \emph{Facebook}, and \emph{Foursquare}, achieving a significant $17$\% increase in performance compared to unimodal learning. Going further, \citet{Factoriz} proposed a factorization method to model the intra-modal and inter-modal relationships within multimodal data inputs, which proved to be important for the incorporation of multimodal data into user profiling, while~\citet{buraya2018multi} instead leveraged the temporal component of the multimodal data, being the first to apply deep learning methods for multi-view personality profiling. While multimodal data has already been tackled in these works, all of them still lacked multi-source cross-social network data processing~\citep{farseev2017360}, which limits their applicability in the majority of real-world scenarios. Therefore, in this work, we base our personality profiling on the PERS framework~\cite{DBLP:conf/mir/YangFF21} that is able to learn from multi-view data for personality profiling by efficiently leveraging highly varied data from diverse social multimedia sources.

Recommender systems have previously used personality traits in several different ways~\cite{DBLP:journals/corr/abs-2101-12153}. Most of them use automatic personality profiling since questionnaire data can hardly be assumed to be available in a real-world setting. Classical approaches use personality features to define a similarity score (proximity function) between users and use it (perhaps together with standard CF proximity) for recommendations~\cite{7904698,10.1145/3030024.3038287,10.1007/978-3-319-10491-1_13} or add personality features to matrix factorization models in a way similar to SVD++~\cite{10.1007/s11257-016-9172-z,10.1007/978-3-319-03524-6_31,10.1145/2505515.2507834}. Approaches based on deep learning have only recently begun to incorporate personality features, and so far these approaches have not used standard personality types but rather inferred their own personality feature vectors~\cite{8784759,DBLP:journals/ijautcomp/HeZL20,somin}. In this work, we propose a hybrid approach that uses predicted MBTI personality type to inform a deep-learning-based recommender system.









\section{Dataset}\label{sec:data}

Our main contribution in this work is to infer a user's personality with multi-modal data from their timeline and recommend suitable content based on the user's content preferences guided by their personality traits. Therefore, to benchmark our model on real-world data, we need a large-scale dataset of user interaction with historical social media content of various brands; it would also be preferable to cover several different industries in order to minimize industry bias. There exist several datasets which consist of posts from social media networks and personality traits. For example, the \emph{myPersonality} dataset~\citep{Mypersonality} had been widely used in research but is not available anymore, and we know of no other large-scale efforts that would be suitable for our task. 
Thus, we need to build our own dataset; in this section, we outline our data gathering and preprocessing methodology.


\subsection{Data acquisition}

We choose \emph{Twitter} as our main data source since it is one of the most open social media outlets, known to concentrate on the users' self-expression rather than their identity and capture more about public personality intended for the broader public because of its high engagement rate. Inspired by~\cite{Francesco}, we composed a list of official \emph{Twitter} accounts of various brands and collected all of their historical posts. We then collected a list of users who liked the posts and the corresponding users' posts and liked posts in their timeline. As for image content of timeline data and brand posts, we selected image previews for video posts and chose the first image for multiple-image posts. Finally, we filtered out brands with less than 100 posts, kept the most recent 100 posts for every brand, and filtered out users who had only one interaction with a brand's post. The resulting dataset statistics are shown in Table~\ref{tbl:recstats}.

\begin{table*}
\centering
\caption{Dataset statistics.}
\label{tbl:recstats}
\begin{tabular}{|l|l|l|l|l|l|l|l|}
\hline
\textbf{Item}     & \textbf{Brands} & \textbf{Brands Posts} & \textbf{Interaction} & \textbf{Users} & \textbf{User Posts} & \textbf{User Images} & \textbf{Sparsity} \\ \hline
\textbf{Quantity} & 48              & 4800                  & 330545         & 41901          & 6547342        & 1407775         & 99.835\%          \\ \hline
\end{tabular}
\end{table*}

\subsection{Data representation}\label{data}

For all models utilized in this work, we extracted both textual and visual features from the ads as follows.
    
For \emph{textual features}:
    \begin{itemize}
        \item for each of Brand posts, we extracted the \emph{tf-idf} features for every post to form the document-term matrix and then applied latent semantic analysis (LSA)~\cite{https://doi.org/10.1002/aris.1440380105} to reduce the textual feature dimension to 100;
        \item for each user, we concatenated their timeline data into the corresponding user-specific ``documents'' and then extracted the \emph{tf-idf} features and the sentiment feature by LIWC lexicon;
    \end{itemize}

For \emph{visual features}, we have chosen to represent visual data in terms of emotional and sentiment-related concepts. To do that, we used a pretrained visual concept detector model \emph{SentiBank}~\cite{Sentibank} to extract visual features; the model outputs a distribution of $2089$ visual sentiment concepts such as \emph{BeautifulNight}, \emph{HappyFace}, or \emph{ClassicDesign}. For each brand post, we extracted the visual sentiment concept distribution, and to represent a user's visual preferences we extracted the concept distribution of every image in the user's timeline and averaged across the concepts, getting a distribution of user preferences with respect to the concepts.

\section{PersiC framework}\label{sec:method}

This section is dedicated to presenting our \emph{Personality-Driven Content Recommendation} (\persrec) framework for the problem of content recommendation for users. We first present the problem setting, then describe the framework itself, and finally report details of the optimization method we used to train the model.

\subsection{Problem setting}

We denote users by $\mathcal{U}=\{u_1,u_2,\ldots,u_i,\ldots, u_n\}$ and brand posts by $\mathcal{P}=\{p_1,p_2,\ldots,p_i,\ldots, p_m\}$. In this notation, the model's goal is to learn a scoring function $f$ to recommend content for users such that for a post $p_x$ that a user $u$ likes and a post $p_y$ that user $u$ did not interact with we would have $f(u,p_x)>f(u,p_y)$. In this work, given a group of brand posts, we aim to learn a ranking model $f$ to rank the content that the user has not interacted with to indicate which posts have a higher chance that the user will like them according to the user's content preferences. 

\subsection{Proposed Method}

\begin{figure}[!t]\centering
     \includegraphics[scale=0.65]{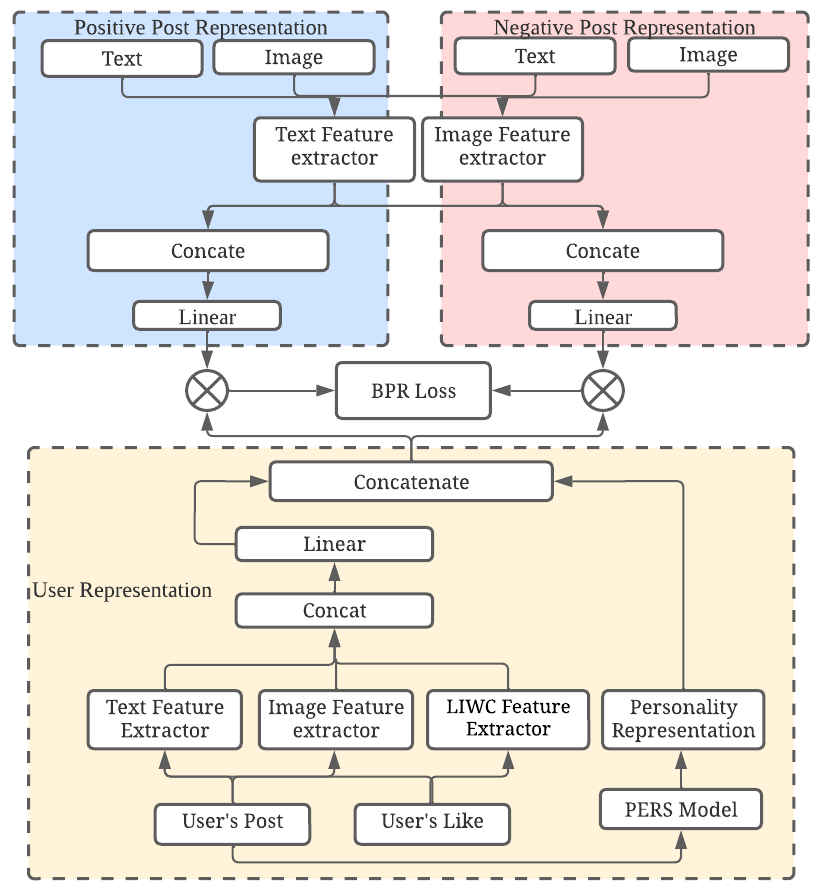}

     \caption{Illustration of the \persrec framework.}
     \label{fig:1}
\end{figure}
The structure of the \persrec framework is illustrated in Figure~\ref{fig:1}. It has two primary components: post representation and user representation learning.

\def\u{\mathbf{u}}

\subsubsection{User Representation Learning.}
The first component is designed to map the users into a common latent space.
The most common approach to learn user representations is through the one-hot representation of features~\cite{onehot1,onehot2} or a fixed size latent vector~\cite{Francesco}. However, in our case personality traits do not provide enough information to distinguish users: with the default approach users who share very similar personality traits would have very high similarity and we would fail to learn fine-grained user representations and fail to recommend relevant content. For this reason, \persrec learns a fine-grained user representation by utilizing the rich multi-modal data from the user's timeline in a social media platform, using personality traits as additional features to inform this representation.

Following the data representation method described in Section~\ref{data}, we represent the timeline data of a user $u_i$ as a collection of features:
$$\mathbf{d}_i=\{T_{pi},I_{pi},T_{li},I_{li},L_{pi},L_{li}\},$$
where $T$ denotes text features, $I$ represents the average value of the image concepts distribution, and $L$ denotes the LIWC feature~\cite{liwc}; the subscript $p$ represents features extracted from the user's timeline and $l$ represents features extracted from the user's historical favorite posts.

To obtain a representation of personality traits, we used the PERS model proposed in~\cite{DBLP:conf/mir/YangFF21}. This model predicts user personality traits from multi-modal inputs (collected user posts), producing four classifiers for each of the four MBTI trait pairs (E-I, S-N, T-F, and J-P). We use the features from the penultimate layer of the PERS model, with $3$ features for each classifier, so in total we obtain a vector $\mathrm{pers}_i \in \mathbb{R}^{12}$.
Using this output, we represent the $i$th user with
$$u_i = \mathrm{Concat}\left(\psi(\mathbf{d}_i), \mathrm{pers}_i\right),$$ where
$$\psi(\mathbf{d}_i) = \mathrm{Linear}\left(\mathrm{Concat}(\mathbf{d}_i)\right)$$
is a linear perceptron with ReLU activation that fuses the previously extracted features $\mathbf{d}_i$; as a result, we get a user latent representation as $\u_i\in\mathbb{R}^{512}$.

\subsubsection{Post Representation}
The second component of \persrec is obtaining post representations. Pure collaborative filtering methods would represent posts as one-hot encoded items and would not utilize any information from the post; our approach follows the latent representation paradigm (see Section~\ref{sec:related} for an overview): \persrec learns from the post's multi-modal data and projects the post in a common latent space with the users, where recommendations will be made.
To achieve this goal, similar to the user representation, we denote the $j$th collection of extracted textual and visual features from a brand's post by $\mathbf{e}_j = \{T_{bj}, I_{bj}\}$,
where $T$ denotes text features, $I$ represents the average value of image concepts distribution, and subscript $b$ denotes features extracted from brand posts. We then add one fully connected layer $\gamma$ with ReLU activations to fuse the textual and image features, and obtain the final representation as
$p_j = \gamma\left(\mathrm{Concat}(\mathbf{e}_j)\right).$

\subsection{Loss function and optimization}

Our dataset $D:=\{(u,p)\}$ consists of user-post pairs that show a post liked by a given user. Inspired by~\cite{Francesco}, we utilize pairwise learning, sampling a fixed number of negative samples for each pair. Based on the assumption that relevant content, i.e., a liked post $p_{\mathrm{pos}}$ should have a higher score than irrelevant content $p_{\mathrm{neg}}$ for a user $u$, we adopted the \emph{Bayesian personalized ranking} (BPR) loss function~\cite{BPR} for our model. Thus, for learning we use $(u, p_{\mathrm{pos}}, p_{\mathrm{neg}})$ triples; to construct them, we take a pair $(u,p_{\mathrm{pos}})\in D$ and uniformly sample $20$ negative posts $p_{\mathrm{neg}}$ (posts that user $u$ did not like) for each positive pair. 

As a result, we learn $f: U \times P\to{\mathbb R}$ by minimizing
$$L(D) = \sum_{u, p_{\mathrm{pos}}, p_{\mathrm{neg}}\in D}\ell(u, p_{\mathrm{pos}}, p_{\mathrm{neg}}),$$
where triples $(u, p_{\mathrm{pos}}, p_{\mathrm{neg}})$ are sampled as above, and
$$\ell(u, p_{\mathrm{pos}}, p_{\mathrm{neg}}) = \log \sigma \left(\hat{y}(u, p_{\mathrm{pos}}) - \hat{y}(u, p_{\mathrm{neg}})\right) - \lambda_{\theta}{{\parallel \Theta \parallel}}^{2},$$
where $\hat{y}(u, p_{\mathrm{pos}})$ and $\hat{y}(u, p_{\mathrm{neg}})$ are scores predicted by our model for user $u$ and items $p_{\mathrm{pos}}$ and $p_{\mathrm{neg}}$ respectively, and $\Theta$ represents the model parameters.

We trained the model with \emph{Adam} optimizer for $30$ epochs, using mini-batches of $64$ user-post pairs shuffled at the beginning of every epoch. We also utilized dropout with rate $0.3$.

\section{Experimental evaluation}\label{sec:eval}

In this section, we report on our experimental study conducted to evaluate the performance of \persrec. We begin by describing the experimental setup, then provide the results of various tests to illustrate the efficiency of our proposed model, including comparisons with baselines and an ablation study to illustrate the impact of personality traits on performance. Finally, we present qualitative case studies on different personality traits.

\subsection{Quality metrics}


In order to evaluate the impact of human personality traits on content recommendation, we have used the following standard metrics:
\begin{itemize}
    \item \emph{Area Under Curve (AUC)}: AUC computes the area under the ROC (Receiver Operating Characteristic) curve for classification problems; larger AUC is better;
    \item \emph{Normalized Discounted Cumulative Gain} (nDCG): nDCG measures the quality of ranking models based on evaluating a ranked list of $n$ top results predicted by the model as follows:
    $$\mathrm{DCG}_{n} =\sum_{i=1}^n \frac{r_i}{\log_2\left(i+1\right)},\quad \mathrm{nDCG}_{n} = \frac{\mathrm{DCG}_{n}}{\mathrm{IDCG}_{n}},$$
    where $r_i$ is the ground truth relevance of the $i$th item on the list, and IDCG is the ideal DCG, i.e., DCG that would be obtained if the results were ranked according to their actual relevance, so $\mathrm{nDCG}_{n}$ is a number between $0$ and $1$; higher nDCG is better;
    \item \emph{$F_{\text{1}}$-measure}: $F_{\text{1}}$-measure is defined as the harmonic average of precision and recall,
    $F_{\text{1}} = \frac{2\cdot\mathrm{precision}\cdot\mathrm{recall}}{\mathrm{precision} + \mathrm{recall}}$;
    higher precision and recall are better, so higher $F_{\text{1}}$-measure is also better.
\end{itemize}

In this work, we truncated the nDCG and $F_{\text{1}}$-measure ranked lists of results at 10 and 50 respectively.

\subsection{Baselines}

To the best of our knowledge, there are no preexisting models developed specifically for personality-driven content recommendation based on brand posts on social media platforms. Thus, we compare our approach with several recommender system baselines (see Section~\ref{sec:related} for a general survey).

\def\p{\mathbf{p}}
\def\q{\mathbf{q}}
\def\x{\mathbf{x}}
\def\w{\mathbf{w}}
\def\v{\mathbf{v}}
\def\bb{\mathbf{b}}
\def\bmu{\boldsymbol{\mu}}
\def\btheta{\boldsymbol{\theta}}
\def\bphi{\boldsymbol{\phi}}
\def\bpsi{\boldsymbol{\psi}}
\def\bsigma{\boldsymbol{\sigma}}
\def\RR{{\mathbb R}}

\subsubsection{Factorization Machines (FM)} Since we are in the implicit feedback setting, for a classical low-rank factorization approach to collaborative filtering we chose factorization machines that show good performance in this setting~\cite{5694074}. For each user $i$ and each item $j$, degree two FM models their possible interaction with a vector $\x\in\RR^n$ that contains one-hot representations of $i$ and $j$ and any additional features. Then the model learns to predict the target variable $y$ (``click''/``no click'' in our case) as
$${\hat y}_{\mathrm{FM}}(\x) = w_0 + \w^\top\x + \sum_{k=1}^n\sum_{l=k+1}^d (\v_k^\top\v_l)w_{kl},$$
where $\w$ and $W$ are weights of the model (in particular, the matrix $W$ represents weights of interactions between features), and $\v_k\in\RR^d$ are feature embeddings, so the user and item embeddings $\p_i\in\RR^d$ and $\q_j\in\RR^d$ in FM are $\v_k$ for the corresponding components of $\x$; see~\cite{5694074} for details about optimization in FM.
% ...

\subsubsection{Neural Collaborative Filtering (NeuCF)} This is a popular approach to recommender systems that generalizes matrix factorization to nonlinear mappings by replacing the inner product with a neural architecture that can learn a more expressive function of data to produce user and item features. We use the neural matrix factorization model from~\cite{10.1145/3038912.3052569} designed for implicit feedback: one-hot representations of each user $i$ and item $j$ are used as input for two different embeddings. Matrix factorization user and item vectors $\p^{\mathrm{MF}}_i\in\RR^d$ and $\q^{\mathrm{MF}}_j\in\RR^d$ are combined into $\bphi^{\mathrm{MF}}_{ij} = \p^{\mathrm{MF}}_i \circ \q^{\mathrm{MF}}_j$ with componentwise multiplication $\circ$; this could lead to the generalized matrix factorization
${\hat y}_{\mathrm{GMF}} = h\left(\w^\top(\p^{\mathrm{MF}}_i \circ \q^{\mathrm{MF}}_j)\right)$
with a weight wector $\w$ and activation function $h$, so for $\w=\mathbf{1}$ and $h=\mathrm{id}$ GMF degenerates into regular matrix factorization. The multilayer perceptron user and item vectors $\p^{\mathrm{MLP}}_i\in\RR^d$ and $\q^{\mathrm{MLP}}_j\in\RR^d$ go through a neural architecture (several MLP layers) to obtain $\bphi^{\mathrm{MLP}}_{ij}$, and then the last layer uses the concatenation of $\bphi^{\mathrm{MF}}_{ij}$ and $\bphi^{\mathrm{MLP}}_{ij}$ to predict $\hat y$; see~\cite{10.1145/3038912.3052569} for details.

\subsubsection{Bilateral Variational Autoencoder for Collaborative Filtering (BiVAE-CF)} This model, presented in~\cite{10.1145/3437963.3441759} learns embeddings via variational autoencoders. This is a generative model that uses the user-item matrix $X$ to learn latent representations $\p_i\in\RR^d$ and $\q_j\in\RR^d$ as follows:
\begin{itemize}
    \item standard Gaussian priors: $p(\p_i) = \mathcal{N}(\mathbf{0}, \mathbf{1})$, $p(\q_j) = \mathcal{N}(\mathbf{0}, \mathbf{1})$;
    \item conditional on $\p_i$ and $\q_j$, the observations in $X$ are drawn from a distribution from the univariate exponential family
    $$p(x_{ij}\mid \p_i,\q_j) = h(x_{ij})e^{\eta(\p_i,\q_j,\omega)x_{ij} - a(\eta(\p_i,\q_j,\omega))};$$
    in our case, we used the Bernoulli distribution since the matrix $X$ is binary (likes);
    \item the (untractable) posterior $p(P,Q\mid X)$ in this model is approximated with a tractable distribution
    $$q(P,Q\mid X) = \left(\prod_iq(\p_i\mid X_{i\ast})\right)\left(\prod_jq(\q_j\mid X_{\ast j})\right),$$
    where
    $q(\p_i\mid X_{i\ast}) = \mathcal{N}(\p_i\mid \bmu(X_{i\ast},\bpsi), \bsigma(X_{i\ast},\bpsi))$ and 
    $q(\q_j\mid X_{\ast j}) = \mathcal{N}(\q_j\mid \bmu(X_{\ast j},\bphi), \bsigma(X_{\ast j},\bphi))$
    are Gaussians whose parameters are functions of the corresponding rows and columns of $X$ with additional variational parameters $\bpsi$ and $\bphi$; in BiVAE-CF, these functions are defined by neural networks with parameters $\bpsi$ and $\bphi$, specifically multilayer perceptrons;
    \item variational parameters $\bpsi$ and $\bphi$ are optimized by minimizing the variational lower bound
    \begin{multline*}
    L_{\mathrm{VAE}} = \sum_{i,j}\mathbb{E}_{q}\left[\log p(x_{ij}\mid \p_i, \q_j)\right] - \\ - \sum_i\mathrm{KL}(q(\p_i\mid X_{i\ast})\| p(\p_i)) - \sum_j\mathrm{KL}(q(\q_j\mid X_{\ast j})\| p(\q_j))
    \end{multline*}
    via the reparametrization trick~\cite{DBLP:journals/corr/KingmaW13} and alternating optimization with respect to user and item variational parameters (we refer to~\cite{10.1145/3437963.3441759} for details).
\end{itemize}
We consider BiVAE-CF to be a strong state-of-the-art baseline for collaborative filtering in our setting.

\subsubsection{Personalized Content Discovery (PCD)} Introduced in~\cite{Francesco}, PCD is the model which is nearest to ours in the overall setting and data used; in particular, it is able to leverage multi-view data and make full use of the dataset we have collected. PCD is also inspired by matrix factorization but designed to learn latent representations for brands and posts on social networks. It proceeds as follows:
\begin{itemize}
    \item one-hot brand id representation and brand associations modeled as a matrix of association vectors $A\in \RR^{n\times k}$ are combined into a brand representation vector
    $\x_b = \sum_{i=1}^n A_{i\ast}\circ \w_b,$
    where $\w_b$ are importance weights for brand $b$, so a brand is represented as a weighted combination of associations;
    \item a post with an image is processed first via a pretrained feature extractor network and then with additional two linear layers with leaky ReLU activations to produce a post feature vector $\x_p$ for every post $p$;
    \item finally, the network parameters are trained with a pairwise ranking loss
    $$L_{\mathrm{PCD}} = \max\left(0, f(b, p_{\mathrm{pos}}) - f(b, p_{\mathrm{pos}}) + \eta \right) + \alpha\sum_{b}|\w_b| + \beta\|\btheta\|_2,$$
    where $f(b, p) = \frac{\x_b^\top\x_p}{\|\x_b\| \|\x_p\|}$ is the normalized scalar product, and the $L_1$ regularizer encourages sparsity in the attention weights $\w_b$.
\end{itemize}
As a result, PCD learns latent brand and post representations together with a fine-grained structure of brand associations. To apply PCD to recommendations with implicit feedback, we replace brands with users, and the rest of the model is unchanged.


\subsection{Evaluation results}

In this section, we evaluate the performance of \persrec against other baselines. We split the dataset into training and test subsets in the $80:20$ ratio, stratified by the number of each user's posts. For a fair comparison, we have reproduced all the baselines with the settings suggested in the original papers and optimized on the training set, then evaluated the performance with the test set separately. Results are listed in the Table~\ref{baseline}.

\begin{table}[!t]
\caption{Experimental evaluation: performance of \persrec and baseline models.}

\label{baseline}
\begin{tabular}{cccccc}
\hline
        & AUC            & $\mathrm{nDCG}_{10}$        & $\mathrm{nDCG}_{50}$        & $\mathrm{F}1_{10}$          & $\mathrm{F}1_{50}$          \\ \hline
MF      & 0.801         & 0.011         & 0.027         & 0              & 0.006          \\
NeuCF   & 0.807          & 0.061          & 0.105          & 0.039         & 0.065         \\
BiVAECF & 0.852           & 0.075           & 0.115          & 0.045           & 0.048         \\
PCD     & 0.881           & 0.082          & 0.121           & 0.048          & 0.091           \\
\persrec & \textbf{0.905} & \textbf{0.092} & \textbf{0.125} & \textbf{0.052} & \textbf{0.095} \\ \hline
\end{tabular}

\end{table}

The table clearly shows that \persrec performs best against other baselines in terms of all considered metrics. The baseline models themselves also perform more or less as expected: the simplest {MF} model is the worst, neural collaborative filtering in the form of \emph{NeuCF} and {BiVAE-CF} performed better than {MF}, but, naturally, they are still inferior when compared to the content-based approaches of PCD and \persrec. Note the very low performance of {MF} that indicates that the recommendation problem we consider is quite hard. These findings indicate that in this sparse multi-view scenario, straightforward collaborative filtering is insufficient, and additional information in the form of rich multi-modal data from both users and items can significantly improve the performance. 

Moreover, the proposed \persrec model has significantly improved results compared to a different commonly used content-based approach, the {PCD} model, across all metrics and especially for the {nDCG} metric where \persrec performed 12\% better. We attribute this finding to two reasons: first, in the method of learning the user representation {PCD} only maps a user id into a fixed sized vector in pre-allocated latent space, while \persrec learns a fine-grained user representation by leveraging rich multi-modal data in the user's timeline and records of their favorite posts, and second and most important, \persrec makes use of the personality traits inferred from the user's social media platform activities. This gives a positive answer to our \textbf{RQ1}:
indeed, even automatically inferred personality traits can significantly improve the performance of downstream recommendation models.
Overall, Table~\ref{baseline} shows that \persrec is able to adapt to the multi-view sparse environment by learning fine-grained multimodal post and user representations and provide recommendations superior to other approaches.

\subsection{Ablation study and the influence of personality traits}

We have conducted an ablation study to evaluate the importance of various features for the user representation and their influence on the final results in terms of recommendation quality metrics. Table~\ref{ablation} shows the results of this ablation study.

First, obviously, the simple one-hot encoding obtains the worst results since it does not utilize additional user information at all.
Learning user representations only from historical post data achieved significantly better scores. Comparing to the performance of PCD in Table~\ref{baseline}, we also see that the performance is improved by leveraging the user's historical data.

Table~\ref{ablation} shows that posts are more useful than likes for the quality of downstream recommendations; note also that the performance generally improves only a little compared to the one-hot encoding if we learn the user representation by the user's historical favorite posts. This can be explained by the high noise in users' historical favorite posts and may be impacted by the algorithm of content recommendation in the social media platform.
Finally, comparing the performance of ``Posts+Likes+Pers'' with ``Posts+Likes'' and ``Posts+Pers'' with ``Posts'', we see significant performance boosts from adding user personality features.
This finding answers positively our \textbf{RQ2}:
introducing personality features into a content recommendation system has been able to substantially improve personalized content recommendation performance. 

\begin{table}[!t]\centering\setlength{\tabcolsep}{3pt}
\caption{Ablation study of various feature combinations for user representations.}

\label{ablation}
\begin{tabular}{cccccc}
\hline
        & AUC            & $\mathrm{nDCG}_{10}$        & $\mathrm{nDCG}_{50}$        & $\mathrm{F}1_{10}$          & $\mathrm{F}1_{50}$          \\ \hline
One-hot         & 0.765           & 0.042                       & 0.079                        & 0.013          & 0.023         \\
Posts           & 0.889          & 0.087                       & 0.113                       & 0.045          & 0.089          \\
Likes           & 0.791          & 0.077                       & 0.095                        & 0.031          & 0.076          \\
Posts+Likes      & 0.891          & 0.087                       & 0.113                        & 0.045          & 0.091           \\
Posts+Pers      & 0.897          & 0.089                       & 0.123                        & 0.047          & 0.091           \\
Posts+Likes+Pers & \textbf{0.905} & \textbf{0.092}              & \textbf{0.125}              & \textbf{0.052} & \textbf{0.095} \\ \hline
\end{tabular}

\end{table}

\begin{figure}[!t]\centering
     \includegraphics[width=\linewidth]{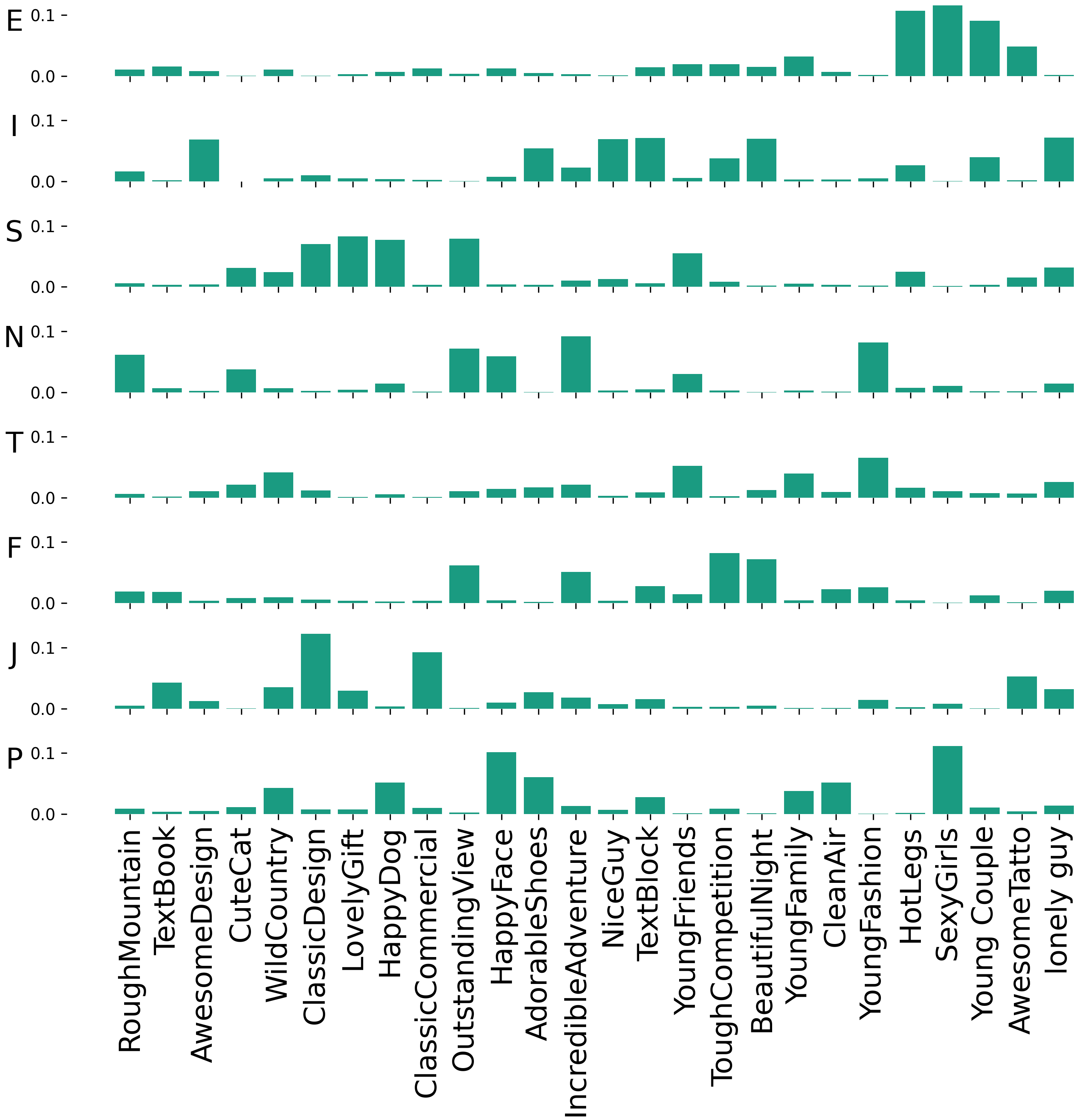}
     \caption{Most correlated concepts for each personality trait from the \emph{Brand}'s timeline data.}
     \label{fig:concept}
\end{figure}

\begin{figure}[!t]\centering
     \includegraphics[width=\linewidth]{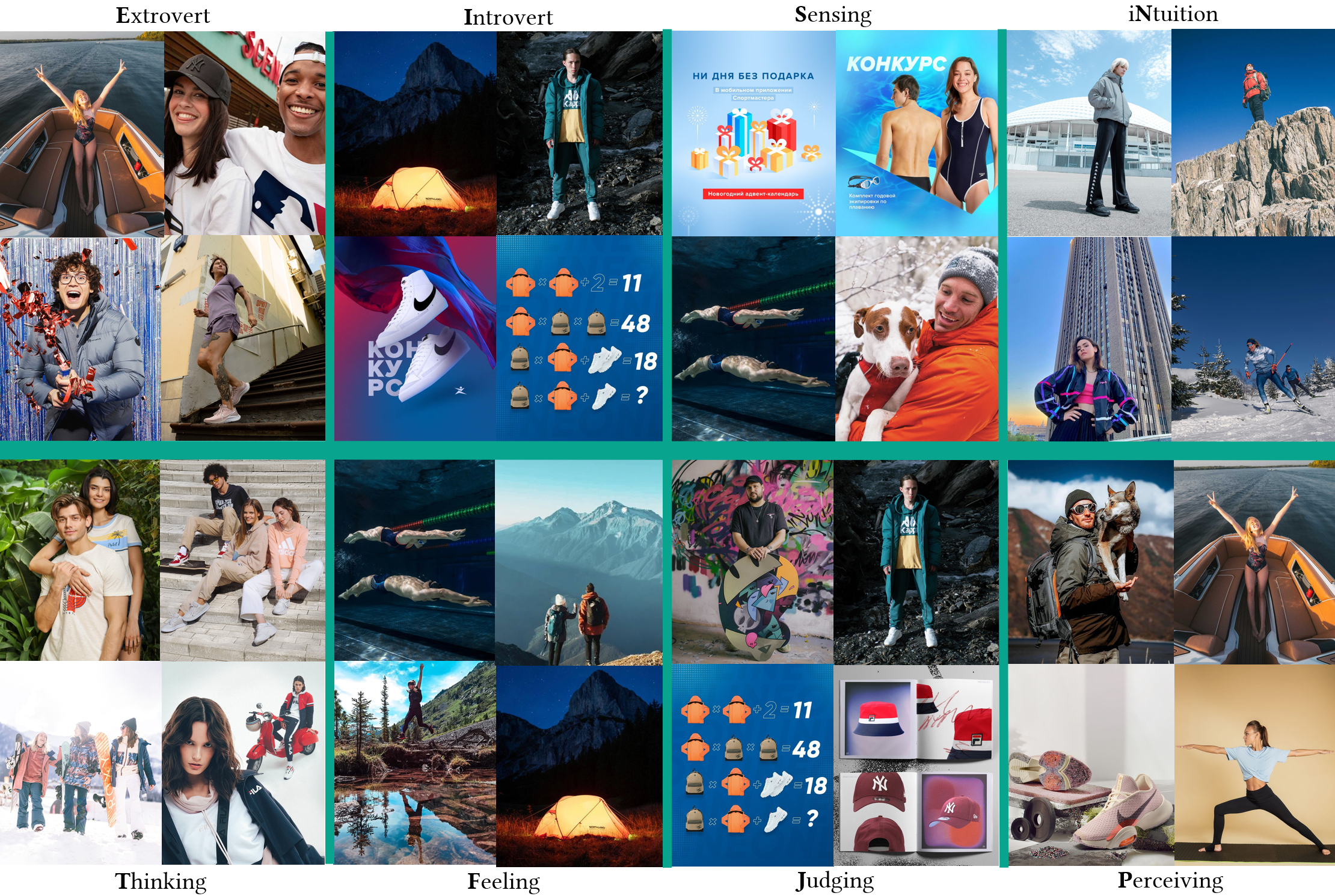}
     \caption{Most highly correlated images for each personality trait from the \emph{Brand}'s timeline data.}

     \label{fig:brand}
     
\end{figure}

\subsection{Qualitative results and case study}
To further evaluate the performance of \persrec, we have conducted a case study with real-world data collected from the audience of a sports brand (hereinafter called \emph{Brand}). \emph{Brand} is a major Decathlon Competitor in one of the European markets. The main goal of the \emph{Brand} is to increase the engagement rate and return on advertising spend (ROAS) by designing better content and choosing proper psychographics.
We have collected historical posts from \emph{Brand}'s and a potential competitor's timelines on \emph{Twitter} and \emph{Instagram}. We then further harvested timeline data from the audiences that had interacted with \emph{Brand}'s posts. Next, as described in Section~\ref{sec:method}, we extracted visual concepts from \emph{Brand}'s and competitor's timelines. Finally, we extracted audience-specific concepts and inferred psychographic attributes from the audiences' timelines.

We begin with a qualitative picture of \persrec's results. Figure~\ref{fig:concept} shows image concepts mined by \emph{SentiBank} that have turned out to be most highly correlated with the $16$ MBTI personality categories based on images from \emph{Brand}'s timeline, and Fig.~\ref{fig:brand} shows several sample images for each personality trait. Note that the distributions of concepts are very different across personality traits; in particular, there are several strong positive correlations: \emph{HotLegs}, \emph{SexyGirls}, \emph{YoungCouple}, \emph{AwesomeTatto}, and \emph{YoungFriends} for the \emph{Extrovert} trait and \emph{Text Block}, \emph{Beautiful Night}, \emph{Nice Guy}, \emph{Awesome Design}, and \emph{LonelyGuy} with the \emph{Introvert} trait. This observation conforms well with the theory that extroverts are more active and enjoy social interaction while introverts are thought-oriented and enjoy spending time alone~\cite{myers1998mbti}.
We also note similarities and differences in concept distributions between traits. For example, \emph{Introvert} and \emph{Judging} are significantly positively correlated to similar concepts such as text and design, reflecting the fact that both traits are thought-oriented and prefer structured, firm decisions~\cite{myers1998mbti}. We can also verify this observation on Fig.~\ref{fig:brand}: \emph{Introvert} and \emph{Judging} traits share similar visual content preferences of numbers, deals, and similar shapes. Fig.~\ref{fig:concept} can yield many such observations: e.g., \emph{SexyGirls} are most important for \emph{Extrovert} and \emph{Perceiving} traits, while \emph{NiceGuy} is best suited for \emph{Introvert} and \emph{Thinking} traits.

Overall, we have found that the distributions of concepts are reasonable and conform well to the assumptions behind this study. 

In this practical case study for \emph{Brand}, we have gone beyond just making recommendations with \persrec into qualitative content generation insights. We have extracted from \persrec the learned mappings and correlations between user content preferences with \emph{Brand}'s post concepts and the common concepts for each personality trait (cf. Figure~\ref{fig:concept}). Then, we have studied insights from the concepts (cf. Figure~\ref{fig:brand}) to guide \emph{Brand}'s content design.

Based on these psychographic insights, with the help of Somin.ai platform, we have automatically created an advertising campaign for the \emph{Brand} on \emph{Meta} (a.k.a. \emph{Facebook Ads}) that leverages predicted personality traits and other user interests~\cite{buraya2017towards}. The above has helped to achieve more than 4.2x cost reduction (for cost per result) compared to the traditional approach adopted by the \emph{Brand}'s agency. This growth in conversion rate has further led to 6.11x more app installs (the advertising budget was increased proportionally) while bringing more insights into the audience's psychographic behavioural traits. 

To be precise, we were able to correlate back the best ad units and the corresponding user personality traits, which brought another insight for the \emph{Brand}: it turned out that purchasing decisions of the majority of the \emph{Brand}'s e-commerce customers are shifted towards long-term implications of owning a product rather than particular product features or its price. As a result, \emph{Brand} has modified the content of their reach and awareness ads that were previously focused on feature- and price-centric content and achieved further improvement of their digital advertising campaign strategy. New fresh audiences visiting the \emph{Brand}'s website were receiving a consistent message on the ``last mile'' of their purchasing journey, just as they did at the stage of familiarizing themselves with the product.

\section{Conclusion}\label{sec:concl}
In this work, we have presented \persrec, a novel personality-driven multi-view content recommender system, which is driven by personality traits inferred from user activities on social media. \persrec is able to capture fine-grained user representations by extracting multi-view features from the user's posts to provide personalized recommendations. Moreover, we have shown a case study with a very successful real-life advertising campaign (improving ROAS by $>4$x and app installs by $>6$x) guided by insights learned by \persrec. Finally, we are also publishing our multi-view large-scale content recommendation dataset for further research in this exciting direction.

For further work, we note that although the \persrec framework improves recommendation performance according to user content preferences, there is an important potential issue that can pollute the results: user activities on their timeline are impacted by content recommendation algorithms adopted by social media platforms such as \emph{Twitter} or \emph{Facebook}. Therefore, we see the further work's backbone direction in studying what is the impact of recommendation algorithms from social media platforms on user activities and how it can be accounted for in recommender systems.

\section{Acknowledgement}
This work was funded by the Russian Science Foundation grant №. 22-11-00135 https://rscf.ru/en/project/22-11-00135/.

\bibliographystyle{ACM-Reference-Format}
\bibliography{new}

\end{document}